\documentclass[a4paper,11pt]{article}
\usepackage{jcappub} % for details on the use of the package, please
\usepackage{graphicx} 
\usepackage{latexsym} 
\usepackage{amssymb}
\usepackage{amsmath} 

\title{\boldmath Spherically symmetric conformal gravity and ``gravitational bubbles''}
\author[a]{V. A. Berezin,}
\author[a,b,1]{V. I. Dokuchaev\note{Corresponding author.}}
\author[a]{and Yu. N. Eroshenko}
\affiliation[a]{Institute for Nuclear Research, Russian Academy of Sciences, \\
	60th October Anniversary Prospect 7a, 117312 Moscow, Russia}
\affiliation[b]{Department of Elementary Particle Physics, National Research Nuclear University MEPhI (Moscow Engineering Physics Institute), Kashirskoye shosse 31, Moscow, 115409, Russia}
\emailAdd{berezin@inr.ac.ru}
\emailAdd{dokuchaev@inr.ac.ru}
\emailAdd{eroshenko@inr.ac.ru}

\abstract{The general structure of the spherically symmetric solutions in the Weyl conformal gravity is described. The corresponding Bach equations are derived for the special type of metrics, which can be considered as the representative of the general class. The complete set of the pure vacuum solutions is found. It consists of two classes. The first one contains the solutions with constant two-dimensional curvature scalar of our specific metrics, and the representatives are the famous Robertson--Walker metrics. One of them we called the ``gravitational bubbles'', which is compact and with zero Weyl tensor. Thus, we obtained the pure vacuum curved space-times (without any material sources, including the cosmological constant) what is absolutely impossible in General Relativity. Such a phenomenon makes it easier to create the universe from ``nothing''. The second class consists of the solutions with varying curvature scalar. We found its representative as the one-parameter family. It appears that it can be conformally covered by the thee-parameter Mannheim--Kazanas solution. We also investigated the general structure of the energy-momentum tensor in the spherical conformal gravity and constructed the vectorial equation that reveals clearly some features of non-vacuum solutions. Two of them are explicitly written, namely, the metrics \`a la Vaidya, and the electrovacuum space-time metrics.}

\begin{document}
\maketitle
\flushbottom

\section{Introduction}
The history of the conformal gravity began in 1918 with the paper by H.~Weyl \cite{Weyl}. His motivation was to construct the unified theory of two (known at the time) fundamental fields: electromagnetic and gravitational ones. Since the electromagnetic field (``identified with the Maxwell equations'') is invariant under conformal transformations, H.~Weyl proposed the conformally invariant Lagrangian for the gravitational field. Then, it was recognized that the Weyl's gravity allows only the massless particles to exist. Moreover the Weyl's Lagrangian was quadratic in Riemann curvature tensor (and its convolutions, the Ricci tensor and scalar curvature).

Meanwhile the General Relativity, based on the Einstein-Hilbert action, which is linear in the scalar curvature, appeared to be the simplest generalization of the nonrelativistic Poisson equation of the Newton's gravity. Because the General Relativity passed successfully several tests and confirmed its validity in the Solar System, it seems needless to deal with a much more sophisticated fourth order partial differential equations, followed from the quadratic Lagrangian.

The interest to conformal gravity was renewed in 1989 by P.~D.~Mannheim and D.~Kaza\-nas \cite{ManKaz89}. They obtained the static spherically symmetric solution, depending on three arbitrary constants, which generalizes the Schwarzschild solution of the General Relativity. The new feature  of this solution is the appearance of the term, linear in radius. It is the latter that may be helpful in understanding the rotation curves of galaxies without introducing the dark matter \cite{ManOBrien12,ManOBrien12b}. Very soon P.~D.~Mannheim \cite{Man90} showed that the conformal gravity vacuum equations allow the de Sitter solution without the cosmological constant. P.~D.~Mannheim found also many other very interesting applications of the conformal gravity (see, e.\,g., reviews \cite{Man07,Man12} and references therein). The causal structure of the static spherically symmetric vacuum solution in conformal gravity was investigated in \cite{EdePar99}. A specific modification of the Weyl conformal gravity was proposed recently by J.~T.~Wheeler \cite{Whe14} by taking into account the variations of all Cartan connection fields instead of the usual fourth-order, metric-only variations. The resulting vacuum field equations reduce to the vacuum Einstein equations, which may be suitable for a renormalizable quantization of the General Relativity.  Cosmological theories with conformal symmetries attract wide attention and have many promising perspectives in quantum field theory \cite{Rubakov}.

We mentioned above that the equations of motion in the conformal gravity contain fourth order derivatives of the metric tensor. This leads to the instabilities of the solutions in classical theory and to the appearance of ghosts in its quantum counterpart. As was mentioned by G.'t Hooft \cite{tHooft14}, the main objection that this theory allows only traceless energy-momentum tensor for the matter fields, can nowadays be removed by making use of the Higgs mechanism for the spontaneous breaking conformal symmetry which was unknown at the times of Einstein and Weyl. Moreover, G.'t Hooft managed to show \cite{tHooft15} that the conventional General Relativity can be reformulated in the conformally invariant form, and the renormalizability condition will require the appearance of the quantum counter-term in the total action which is exactly the same as that of the conformal gravity. In this way the desired stability of the solutions may also be reached. 

Meantime, our motivation for studying the conformal gravity is twofold. First, we are interested in phenomenological description of particle creation in the presence of external gravitational and electromagnetic fields \cite{Ber87,Ber14}. As was shown by Ya.~B.~Zel'dovich, I.~D.~Novikov and A.~ A.~Starobinsky \cite{ZeldNovStar1974}, the rate of particle creation in (weakly anisotropic) cosmological models is proportional to the square of the Weyl tensor. In our scheme it will enter into the total action integral with the Lagrange multiplier. Combined with the conformal gravity action, we arrive at the very curious model with the effectively varying coupling constant. The simplest way to probe such a feature is, of course, the homogeneous and isotropic cosmological model. However, to our surprise, it appeared that for  all these models the Weyl tensor vanishes identically. So, in the conformal gravity the isotropic and  homogeneous cosmological can be only the vacuum space-time. And this is our second motivation reason.

The vacuum space-time is a very good candidate for the creation of the universe ``from nothing'' \cite{Vil82}. We called ``the gravitational bubble'' one of the vacuum solutions in the conformal gravity, which is compact and has zero Weyl tensor.

The idea that the initial state of the universe should be conformal invariant is advocated also by R.~Penrose \cite{Penr10,Penr14} and G.~'t~Hooft \cite{tHooft14}. But, how such an empty universe could be filled with the matter (necessary massless in conformal gravity)? Surely, due to quantum fluctuations, which will cause the local deviations of the Weyl tensor from zero. This, in turn, will cause the creation of the massless particles, which, moving with the speed of light, will deviate the Weyl tensor from zero in much more wider region, and so on, and so forth. This process reminds the detonation wave and looks like the burning of the vacuum \cite{BerKuzTk83}.

The questions arise: for how long this could last and what could be the result? In 1970 Ya.B.Zel'dovich showed \cite{Zeld70} (see also \cite{ZeldStar71,LukashStar74}) that the particle creation process can make the universe isotropic, if its evolution started from the highly anisotropic (homogeneous) Kazner-type initial state. Our situation is somewhat similar. At the same time, now we encounter the problem. The isotropization  and homogenization will lead to decreasing of both the particle creation rate and the square of the Weyl tensor. This is exactly the desirable behavior in our scheme. However, the problem is in the absence of the isotropic and homogeneous solution in the conformal gravity, except the vacuum one. This controversy is overcome by taking into account not only the energy -momentum tensor of the already created particle, but  also the quantum corrections coming from the vacuum polarization and trace anomaly \cite{ZeldStar71,Parker69,GribMam69,ZeldPit71}.

In one-loop approximation the corresponding counter-terms can be incorporated into the (gravitational) action integral as the curvature scalar (manifesting the renormalization of the Newton's constant), the cosmological constant and the the term, quadratic in the Ricci tensor.

In our phenomenological scheme all this is encoded in the particle creation law, accompanied by the Lagrange multiplier. Because from the very beginning there is no Newton's constant, its renormalization means the emergency. Thus, while the Weyl tensor is gradually vanishing, the counter-terms, linear in the Ricci tensor and scalar curvature is becoming dominant. In result, the General Relativity comes into play. The irreversible particle creation process is the mechanism for the inevitable spontaneous symmetry breaking. We see that the instabilities, inherent in the theories with higher (than the second) derivatives, play here the very positive role, like in the Starobinsky's inflationary scenario \cite{Star80}

The paper is organized as follows: In Section~\ref{SphSymm} we derive the Bach equation in the spherical symmetry by using the $2+2$ decomposition of the metric. In Section~\ref{VacSol}  we define the general structure of the vacuum solutions in conformal gravity by using the double null coordinates. In Section~\ref{Restore} we describe the restoration of radial coordinate in the derived spherically symmetric vacuum solution and compare it with the known solution by P.~D.~Mannheim, D.~Kazanas \cite{ManKaz89}. In Section~\ref{ConstR} we describe the structure of vacuum cosmological solutions with a constant curvature by using the general solution of the Liouville equation. In Section~\ref{EnMomTensor} we consider the properties of the emergent energy-momentum tensor in Bach equation. In Section~\ref{Vaidya} we consider the general properties \`a la Vaidya solution in the conformal gravity. Finally in Conclusion we briefly summarize the results.

\section{Bach equation in spherical symmetry}
\label{SphSymm}

The 4D curvature-quadratic action of the Weyl conformal gravity is
\begin{equation}
S=-\alpha_0\int C^{\mu\nu\lambda\sigma}C_{\mu\nu\lambda\sigma}\sqrt{-g}d^4x,
\label{C2}
\end{equation}
where the Weyl tensor is defined as
\begin{eqnarray}
&&C_{\mu\nu\lambda\sigma}=R_{\mu\nu\lambda\sigma}
+\frac{1}{2}(R_{\mu\sigma}g_{\nu\lambda}+R_{\nu\lambda}g_{\mu\sigma}-
\nonumber \\
&&-R_{\mu\lambda}g_{\nu\sigma}-R_{\nu\sigma}g_{\mu\lambda})+\frac{1}{6}R(g_{\mu\lambda}g_{\nu\sigma}-g_{\mu\sigma}g_{\nu\lambda})
\label{Weyltensor}
\end{eqnarray}
and $\alpha_0=const$. The corresponding fourth order dynamical equations for the conformal gravity were derived by R. Bach \cite{Bach21}:
\begin{equation}
C^{\mu\sigma\nu\lambda}_{\phantom{0}\phantom{0}\phantom{0}\phantom{0};\lambda;\sigma}
+\frac{1}{2}C^{\mu\lambda\nu\sigma}R_{\lambda\sigma}=\frac{1}{8\alpha_0}\,T^{\mu\nu},
\label{Bach}
\end{equation}
where $T^{\mu\nu}$ is a matter energy-momentum tensor.

Our main goal is to investigate the general properties and structure of the spherically symmetric vacuum solutions of the Bach equations (\ref{Bach}). In \ref{Bacheqapp} we present the sketch for the derivation of the dynamical equations of the conformal gravity from the Weyl action (\ref{C2}) in the general case. In this Section we derive the corresponding form of the Bach equations in the case of the spherical symmetry.

We use the following sign conventions: the metric signature $(+,-,-,-)$, the Riemann curvature tensor
\begin{equation}
R_{\phantom{0}\nu\lambda\sigma}^{\mu} =
\frac{\partial\Gamma_{\nu\sigma}^{\mu}}{\partial x^\lambda}
- \frac{\partial\Gamma_{\nu\lambda}^{\mu}}{\partial x^\sigma}
+ \Gamma_{\varkappa\lambda}^{\mu}\Gamma_{\nu\sigma}^{\varkappa}
- \Gamma_{\varkappa\sigma}^{\mu}\Gamma_{\nu\lambda}^{\varkappa},
\label{Riemann}
\end{equation}
and the Ricci tensor
\begin{equation}
R_{\nu\sigma}= R_{\phantom{0}\nu\mu\sigma}^{\mu}.
\label{Ricci}
\end{equation}
Our formalism is based on the $2+2$ decomposition, and we will use the following agreements for indices:
\begin{itemize}
	\item Greek indices from the middle of the alphabet $\mu,\nu,\sigma,\lambda$ run the values $0,1,2,3$;
	\item Latin indices $i,j,k,l$ run the values $0,1$;
	\item Greek indices from the beginning of the alphabet $\alpha,\beta,\gamma,\delta$ run the values $2,3$. 	
\end{itemize}
We use the following $2+2$ decomposition of the metric in the case of spherical symmetry:
\begin{eqnarray}
ds^2 &=& \gamma_{ik}dx^i dx^k-r^2(x)(d\theta^2+\sin^2\theta d\varphi^2) \nonumber \\
&=& r^2(x)[\tilde\gamma_{ik}dx^i dx^k-(d\theta^2+\sin^2\theta d\varphi^2)]
=r^2(x)[d\tilde s_2^2 -d\Omega^2],
\label{r2s2}
\end{eqnarray}
where
\begin{equation}
d\tilde s_2^2=\tilde\gamma_{ik}dx^i dx^k
\label{2dimflat}
\end{equation}
is a 2-dimensional metric and $d\Omega^2=d\theta^2+\sin^2\theta d\varphi^2$ is a line element on the 2-dimensional unit sphere. Conformal symmetry allows us to withdraw the $r^2$ from the metrics (\ref{r2s2}), therefore in the following we will write the metric in the form
\begin{equation}
ds^2 =\tilde\gamma_{ik}dx^i dx^k-(d\theta^2+\sin^2\theta d\varphi^2).
\end{equation}
Further in this article the semicolon ``$;$'' means the covariant derivatives with respect to the original fourth-dimensional metrics $g_{\mu\nu}$, and the vertical bar ``$|$'' defines the covariant derivatives for the two-dimensional metrics $\gamma_{ik}$.

The corresponding  non-zero affine connections are
\begin{eqnarray}
\Gamma_{kl}^{i}&=&\tilde\Gamma_{kl}^{i}, \\
\Gamma_{\mu\nu}^{2} &=& \frac{1}{2}g_{\mu\nu,2}, \quad
\Gamma_{33}^{2} =-\sin\theta\cos\theta, 
\\
\Gamma_{\mu\nu}^{3} &=& -\frac{1}{2\sin^2\theta}(g_{3\mu,\nu}+g_{3\nu,\mu}), \quad
\Gamma_{23}^{2}=\cot\theta,
\end{eqnarray}
where tilde `$\,\tilde{}\,$' means the two-dimensional space-time with respect to the $\gamma_{ik}$. The non-zero components of the Riemann curvature tensor (\ref{Riemann}) are the following:
\begin{eqnarray}
R_{iklm} &=& \tilde R_{iklm}=\frac{\tilde R}{2}(\gamma_{il}\gamma_{km}-\gamma_{im}\gamma_{kl}),
\\
R_{km} &=& \frac{\tilde R}{2}\gamma_{km}, \quad R_{22}=1, R_{33}=\sin^2\theta,
\\
R_2^2 &=& R_3^3=-1,
\end{eqnarray}
where $\tilde R$ is a two-dimensional scalar curvature.
\begin{eqnarray}
R_{i2\lambda\sigma} &=& \frac{1}{2}(R_{i\sigma}g_{2\lambda}+R_{2\lambda}g_{i\sigma}
-R_{i\lambda}g_{2\sigma}-R_{2\sigma}g_{i\lambda})
 +\frac{1}{6}R(g_{i\lambda}g_{2\sigma}-g_{i\sigma}g_{2\lambda}),
\\
R_{i2k2} &=& \frac{1}{12}(\tilde R-2)\gamma_{ik}, \\
R_{2323} &=& \frac{1}{6}(\tilde R-2)\sin^2\theta,
\\
R_{i3k3} &=& \frac{1}{12}(\tilde R-2)\gamma_{ik}\sin^2\theta.
\end{eqnarray}
And the corresponding four-dimensional scalar curvature is
\begin{equation}
R=\tilde R-2.
\end{equation}
The Weyl tensor (\ref{Weyltensor}) has the following non-zero components.
\begin{equation}
C_{iklm}=\frac{1}{6}(\tilde R-2)(\gamma_{il}\gamma_{km}-\gamma_{im}\gamma_{kl}),
\end{equation}
\begin{equation}
C_{2323}=\frac{(\tilde R-2)\sin^2\theta}{6}, \;\; C_{i2j2}
=\frac{(\tilde R-2)\gamma_{ij}}{12},
\end{equation}
\begin{equation}
C_{i3j3}=C_{i2j2}\sin^2\theta.
\end{equation}
The ``squared'' Weyl tensor $C_{\mu\nu\sigma\lambda}C^{\mu\nu\sigma\lambda}$ in the action (\ref{C2}) can be calculated either directly or by using the identity
\begin{equation}
C_{\mu\nu\sigma\lambda}C^{\mu\nu\sigma\lambda}
=R_{\mu\nu\sigma\lambda}R^{\mu\nu\sigma\lambda}-2R_{\mu\nu}R^{\mu\nu}+\frac{1}{3}R^2.
\end{equation}
In the spherically symmetric case both methods give us
\begin{equation}
C^2 \equiv C_{\mu\nu\lambda\sigma}C^{\mu\nu\lambda\sigma}=\frac{1}{3}(\tilde R-2)^2.
\end{equation}
The corresponding Bach tensor
\begin{equation}
B_{\mu\nu} =
C_{\mu\varkappa\nu\sigma}^{\phantom{0}\phantom{0}\phantom{0}\phantom{0};\sigma\varkappa}
+\frac{1}{2}R^{\varkappa\sigma}C_{\mu\varkappa\nu\sigma}
\end{equation}
has the following non-zero components:
\begin{eqnarray}
\label{Bachnonzero}
B_{ik} &=& \frac{1}{6}(\tilde R_{|p}^{|p}\gamma_{ik}
-\hat R_{|ki})+\frac{\tilde R^2-4}{24}\gamma_{ik},
\\
B_{22} &=& \frac{1}{12}\left[\tilde R_{|p}^{|p}+\frac{1}{2}(\tilde R^2-4)\right],
\\
B_{33} &=& \frac{1}{12}\left[\tilde R_{|p}^{|p}-\frac{1}{2}(\tilde R^2-4)\right]\sin^2\theta
=B_{22}\sin^2\theta.
\end{eqnarray}

\section{General structure of vacuum solutions}
\label{VacSol}

From the results of Section~\ref{SphSymm} one has the following Bach equations in vacuum
\begin{eqnarray}
B_{ik} &=& \frac{1}{6}(\tilde R_{|p}^{|p}\gamma_{ik} \label{Bachik}
-\hat R_{|ki})+\frac{\tilde R^2-4}{24}\gamma_{ik}=0,
\\
B_{22} &=& \frac{1}{12}\left[\tilde R_{|p}^{|p}+\frac{1}{2}(\tilde R^2-4)\right]=0.
\label{Bach22}
\end{eqnarray}
In this section we will present their general solutions.

\subsection{Double null coordinates}

Let us introduce the double null coordinates in the 2-dimensional metrics:
\begin{equation}
ds_2^2=\gamma_{ik}dx^i dx^k=2e^{2\omega}dudv=2Hdudv,
\label{doublenull}
\end{equation}
i.~e.
\begin{equation}
\gamma_{ik}=\left( \begin{array}{cc}
0 & H \\ H & 0
\end{array} \right), \quad
\gamma^{ik}=\left( \begin{array}{cc}
0 & H^{-1} \\  H^{-1} & 0
\end{array} \right).
\end{equation}
The non-zero components in this case
\begin{equation}
\Gamma_{uu}^u=\frac{H_{,u}}{H}, \quad \Gamma_{vv}^v=\frac{H_{,v}}{H},
\end{equation}
\begin{equation}
R_{uvvu}=(H_{,u}H_{,v}-HH_{,uv})/H,
\end{equation}
\begin{equation}
R_{uv}=(H_{,u}H_{,v}-HH_{,uv})/H^2,
\label{ro1t1}
\end{equation}
\begin{equation}
\tilde R=\gamma^{ik}R_{ik}=\frac{2}{H}R_{uv}.
\end{equation}
With the use of (\ref{ro1t1}) we have
\begin{equation} \label{scalar}
\tilde R=\frac{2}{H^3}(-HH_{,uv}+H_{,u}H_{,v})=R+2,
\end{equation}
\begin{equation} \label{scalar2}
C_{1010}=-\frac{1}{6}RH^2, \;\;\; C_{2021}=\frac{1}{6}RH, \;\;\;
C_{3031}=\frac{1}{6}RH\sin^2\theta, \;\;\; C_{3232}=\frac{1}{6}R\sin^2\theta.
\end{equation}
For any scalar $A$ we have
\begin{equation}
A_{|p}^{|p}=\gamma^{lm}A_{|l|m}=\frac{2}{H}A_{|uv}
=\frac{2}{H}(A_{,uv}-\Gamma_{uv}^{\mu}A_{,\mu})=\frac{2}{H}A_{,uv},
\end{equation}
The corresponding Bach equations in vacuum are the following ones:
\begin{eqnarray}
B_{uv} &=& 0 \quad  \Rightarrow \quad \tilde R_{,uv}=\frac{H}{4}(\tilde R^2-4), \label{Buv0}
\\
B_{22} &=& 0 \quad  \Rightarrow \quad \tilde R_{,uv}=\frac{H}{4}(\tilde R^2-4), \label{B220}
\\
B_{uu} &=& 0 \quad  \Rightarrow \quad \tilde R_{,uu}=0, \quad \Rightarrow
\quad \frac{\tilde R_{,uu}}{\tilde R_{,u}}=\frac{H_{,u}}{H}, \label{Buu0}
\\
B_{vv} &=& 0 \quad  \Rightarrow \quad \tilde R_{,vv}=0, \quad \Rightarrow
\quad \frac{\tilde R_{,vv}}{\tilde R_{,v}}=\frac{H_{,v}}{H}. \label{Bvv0}
\end{eqnarray}

\subsection{General solution of the Bach equation in the double null coordinates}

From (\ref{Buu0}) and (\ref{Bvv0}) we have the first integrals:
\begin{equation} \label{c12}
\tilde R_{,u}=c_1(v)H, \quad \tilde R_{,v}=c_2(u)H.
\end{equation}
We choose new null coordinates $\tilde u$ and $\tilde v$ to ``absorb'' the functions $c_1(v)$ and $c_2(u)$ and retain the coordinate ``directions'', i.\,e., to hold true the inequalities $d\tilde u/du>0$ and $d\tilde v/dv>0$. This is possible, in general, if $c_1\neq0$ and $c_2\neq0$, let it be the case. In new coordinates we have $\tilde R_{,\tilde u}=\pm \tilde R_{,\tilde v}$ or $\tilde R (z)=\tilde R(u\pm v)$ and $\tilde R_{,\tilde u\tilde v}=\pm \tilde R_{,\tilde u\tilde u}$. It means that our metrics, Eq.~(\ref{doublenull}), admits the Killing vector, time-like in the case $z=u-v$, and space-like in the case $z=u+v$.

Let us now $c_1(v)=1$, then $\tilde R_{,u}=H$ and
\begin{equation}
\pm\tilde R_{,zz}=\tilde R_{,z}(\tilde R^2-4), \quad z=\tilde u\pm\tilde v.
\end{equation}
We are lowering the order of differential equation by the standard procedure:
\begin{equation}
\tilde R_{,z}\equiv Y(R_{,z}) \quad \Rightarrow
\frac{dY}{d\tilde R}=\frac{\tilde R^2}{4}-1 \quad \Rightarrow
\pm Y=\frac{\tilde R^3}{12}-\tilde R+C_0.
\end{equation}

So, the general spherically symmetric solution of the Bach equation in vacuum has the form
\begin{eqnarray}
ds_2^2&=&\pm \frac{1}{6}(\tilde R^3-12\tilde R+C_0)dudv, \\
\tilde R&=&\tilde R(z), \quad z=u \pm v.
\end{eqnarray}
Another choice, $\tilde R_{,z} = - H \,\, (c_1(v) = - 1)$, gives exactly the same result.

Surely, one may continue and solve the remaining first order differential equation in order to obtain the solutions $\tilde R (u,v)$ in the region between the horizons ( i.\,e., between the roots of the equation $\tilde R_{,z} = 0$). But, we will not do this here. Instead, we. first, introduce the more customary ``time'' and ``space'' coordinates $\eta$ and $\xi$ by the relations $u = \eta - \xi$ and $v = \eta + \xi$. Then, $d u d v = d \eta^2 - d \xi^2$. Remember, that the only metric coefficient $H$ depends either on $\xi$, or on $\eta$,  i.\,e., one of these coordinates is directed along the Killing vector, while the other is orthogonal to it. The next step is to use the two-dimensional curvature scalar $\tilde R$ itself as this second coordinate. In this case the two-dimensional metrics takes very simple and elegant form
\begin{eqnarray}
\label{2-dimfin}
d \tilde s^2_2 &=& A d \eta^2 - \frac{d \tilde R^2}{A} \, ,
\end{eqnarray}
where
\begin{eqnarray}
\label{A}
A &=& \frac{1}{6} \left (\tilde R^3 - 12 \tilde R + C_0  \right)\, .
\end{eqnarray}
If $A > 0$, the coordinate $\tilde R$ is spatial, and if $A < 0$, it plays the role of time. To convert the above two-dimensional metrics into the four-dimensional one, we should just to add the line element of the unit two-dimensional sphere. So,
\begin{eqnarray}
\label{4-dimfin}
d \tilde s^2_4 &=& A d \eta^2 - \frac{d \tilde R^2}{A} - (d \theta^2
+ \sin^2 \theta d \varphi^2) \, ,
\end{eqnarray}
and this metrics can be called ``the representative'' of the whole family of the vacuum solutions with non-zero two-dimensional curvature scalar.

For completeness, let us put now $c_1 (v) = 0$. Then, $\tilde R_{,u}=0$, $\tilde R_{,uv}=0$ and $c_2(u)=0$. The vacuum solutions are $\tilde R=const=\pm2$. The detailed investigation will be described in the separate Section.

\subsection{Restoration of the radial coordinate}
\label{Restore}

The general four-dimensional vacuum solution solution can be written in the form
\begin{equation}
d s_4^2 =\Phi^2(\eta,\tilde R)\left[A d\eta^2-\frac{d\tilde R^2}{A}-(d\theta^2+\sin^2\theta d\varphi^2)\right],
\label{gensol}
\end{equation}
where $\Phi$ plays the role of the physical radius of the sphere. Let us make the coordinate transformation $(\eta, \, \tilde R) \rightarrow (\eta, \, \Phi)$. We prefer to leave the time coordinate unchanged because it reminds us of the Killing vector. The result is
\begin{eqnarray}
\label{4dimraddius}
d s^2_4 &=& \left[\Phi^2 A - \frac{\Phi^2}{A} \frac{(\frac{\partial \Phi}{\partial \eta})^2}{(\frac{\partial \Phi}{\partial \tilde R})^2} \right] d \eta^2 - 2 \frac{\Phi^2}{A} \frac{\frac{\partial \Phi}{\partial \eta}}{\frac{\partial \Phi}{\partial \tilde R}} d \eta d \Phi 
 \nonumber \\
&-& \frac{\Phi^2}{A} \left(\frac{\partial \Phi}{\partial \tilde R}\right)^{\!-2}\! d \Phi^2 - \Phi^2 (d \theta^2 + \sin^2 \theta d \varphi^2) \, .
\end{eqnarray}
We see, that, in general, the metrics is not diagonal. Since the radius has the very obvious physical meaning, we would like to construct yet another representative, with the radius as one of the coordinates, and at the same time, where there exists the Killing vector. It is possible, if the radius depends only on $\tilde R$, we will denote it $r (\tilde R)$. To specify the function $r (\tilde R)$ we demand that the resulting metrics should be the Mannheim--Kazanas solution \cite{ManKaz89}. For this, it is sufficient to require
\begin{equation}
\label{radiusfunction}
\frac{d \tilde R}{d r} = \frac{C_1}{r^2} \quad \Longrightarrow \tilde R = - \frac{C_1}{r} + C_2
\end{equation}
The, after redefinition of the time coordinate, we get
\begin{equation}
\label{ourKM}
d s^2_4 = F d t^2 - \frac{d r^2}{F} - r^2 (d \theta^2 + \sin^2 \theta d \varphi^2)
\end{equation}
For the function F(r) we find the polynomial expression
\begin{equation}
F(r) = \frac{C_2}{2} - \frac{C_1}{6 r} + \frac{(4-C_2^2)}{2 C_1} r + \frac{C_2(C_2^2 - 12)- C_0}{6 C_1^2} r^2
\end{equation}
One can easily express the coefficients in the Mannheim--Kazanas solution in terms of our $C_1,\,C_2,\,C_0$ coefficients by the relations:
\begin{equation}
\frac{C_2}{2}=1-3\beta\gamma, \quad \beta(2-3\beta\gamma)=\frac{C_1}{6},
\end{equation}
\begin{equation}
\gamma=\frac{4-C_2^2}{2C_1}, \quad k=\frac{C_2(12-C_2^2)+C_0}{6C_1^2}.
\end{equation}

It is interesting to compare the obtained vacuum solution with the well known spherically symmetric vacuum solutions to the Einstein equations of General Relativity. The latter ones are described by the metrics of the form
\begin{equation}
ds^2 =Fdt^2-\frac{dr^2}{F}-r^2(d\theta^2+\sin^2\theta d\varphi^2),
\end{equation}
\begin{equation}
F=1-\frac{2Gm}{r}-\frac{\Lambda}{3}r^2,
\end{equation}
where $G$ is the Newton's constant, $m$ is the mass parameter, and $\Lambda$ is the so called cosmological term. When $\Lambda=0$, we have the famous Schwarzschild solution, if $m=0$ --- the de Sitter ($\Lambda>0$) and anti-de Sitter ($\Lambda<0$) solutions, and for nonzero mass they called Schwarzschild-de Sitter and Schwarzschild-anti-de Sitter solutions, correspondingly. In order to reach our goal we need to calculate the scalar curvature $\tilde R$ for the truncated metrics
\begin{equation}
d\tilde s_2^2 =\frac{Fdt^2}{r^2}-\frac{dr^2}{Fr^2},
\end{equation}
and made a transformation to the coordinates $(t,\tilde R)$. The result is
\begin{equation}
\tilde R=2-\frac{12Gm}{r},\label{dash3}
\end{equation}
\begin{equation}
d\tilde s_2^2 =\frac{(12Gm)^2F}{r^2}d\left(\frac{t}{12Gm}\right)^2
-\frac{r^2}{(12Gm)^2F}d\tilde R^2 
=Ad\eta^2-\frac{d\tilde R^2}{A},
\end{equation}
where
\begin{equation}
\eta=\frac{t}{12Gm}; \qquad A=\left(\frac{12Gm}{r}\right)^2F.
\end{equation}
Substituting $r$ from Eq.~(\ref{dash3}) one gets
\begin{equation}
A=\frac{1}{6}[\tilde R^3 -12\tilde R+16-2(12Gm)^2\Lambda].
\end{equation}
We see that the spherically symmetric vacuum solutions of General Relativity are reproduced for the special value of our $C_0$, $C_1$ and $C_2$ constants, namely
\begin{equation}
C_2=0, \quad C_1=12Gm, \quad C_0=16-2(12Gm)^2\Lambda.
\end{equation}

\section{Solutions with constant curvature}
\label{ConstR}

The vacuum solutions with a constant curvature $\tilde R$ require $\tilde R = \pm 2$, but it appeared possible to find all the spherically symmetric solutions with $\tilde R = const$ in conformal gravity. And this is the subject of the present Section. 

The two-dimensional space-time metric (\ref{2-dimfin}) with a constant curvature $\tilde R$ in the double null coordinates $(u,v)$ is defined by the solution of equation (\ref{scalar}):
\begin{equation} \label{Rconst}
\tilde R=\frac{2}{H^3}(-HH_{,uv}+H_{,u}H_{,v})=const, \quad H=e^{2\omega}.
\end{equation}
In terms of the function $\omega$ this equation takes the form
\begin{equation}
\omega_{,uv}=-\frac{\tilde R}{4}e^{2\omega}.
\end{equation}
This is the Liouville equation in the differential geometry (see, e.\,g., \cite{Polyanin}):
\begin{equation} \label{Liouville}
\frac{\partial^2Y}{\partial u \partial v}=ae^{\lambda Y}
\end{equation}
for function $Y=\omega$ with constants $a=-\tilde R/4$ and $\lambda=2$. The general solution Liouville equation is
\begin{equation} \label{solLiouville}
Y=\frac{1}{\lambda}[f(u)+g(v)]
-\frac{2}{\lambda}\ln\left|k\!\int\exp[f(u)]\,du
+\frac{a\lambda}{2k}\int\exp[g(v)]\,dv\right|, 
\end{equation}
where $f=f(u)$, $g=g(v)$ are the arbitrary functions, and $k$ is an arbitrary constant.

The two-dimensional metrics with constant curvature can be written in the form:
\begin{equation}
ds_2^2=\frac{2}{k^2}\frac{dFdG}{\left(F-\frac{\tilde R}{4k^2}G\right)^2}
=\frac{8}{|\tilde R|}\frac{d\tilde ud\tilde v}{(\tilde u\pm\tilde v)^2},
\end{equation}
where
\begin{equation}
F=\int e^{f(u)}du=\tilde u, \quad
G=\int e^{g(v)}dv=\left|\frac{2k}{\tilde R}\right|^2\tilde v.
\end{equation}
The metric function $H$ depends, respectively, on $\tilde u-\tilde v$ at $\tilde R>0$ and $\tilde u+\tilde v$ at $\tilde R<0$. For $H>0$ we have $\tilde u=t-x$ and $\tilde v=t+x$,  i.\,e., every two-dimensional space-time of this kind admits the Killing vector, either time-like, or space-like.

In the case $\tilde R>0$ we have $z=u-v$, $w_{uv}=-w_{zz}$. The Liouville equation takes the form
\begin{equation}
\omega_{,zz}=\frac{\tilde R}{4}e^{2\omega}.
\end{equation}
By defining new function $y=2\omega$, we may reduce the order of the differential equation:
\begin{equation}
y_{,zz}=\frac{\tilde R}{2}e^{y}, \quad y_{,z}=p(y), \quad y_{,zz}=p'p \quad \Rightarrow
p=\pm\sqrt{\tilde Re^{y}+c}.
\end{equation}
Here $c$ is the integration constant. Solution of the last equation is
\begin{equation} \label{Rconst2}
z=\pm\int\frac{dy}{\sqrt{\tilde Re^{y}+c}}
=\mp2\int\frac{d\alpha}{\sqrt{\tilde R+c \alpha^2}}, \quad \alpha=e^{-y/2}.
\end{equation}
In the special case of $c=0$ solution is
\begin{equation}
\frac{\sqrt{\tilde R}}{2}|z|=e^{-y/2}.
\end{equation}
This solution with $v=t+x$, $u-v=-2x$ corresponds to the metric
\begin{equation}
ds^2_2=\frac{2}{\tilde R x^2}(dt^2-dx^2).
\end{equation}
In the case of $c >0$ solution (\ref{Rconst2}) takes the form
\begin{equation} \label{Rconst2b}
z=\mp\frac{2}{\sqrt{c}}{\rm arcsh}\left(\frac{\alpha\sqrt{c}}{\sqrt{\tilde R}}\right)
\end{equation}
with
\begin{equation} \label{Rconst2c}
\alpha=\frac{\sqrt{\tilde R}}{\sqrt{c}}{\rm sh}\left(\mp\sqrt{c}z/2\right)=e^{-y/2}.
\end{equation}
\begin{equation} \label{Rconst2d}
H=e^y=\frac{c}{\tilde R {~\rm sh}^2\left(\sqrt{c}z/2\right)}
=\frac{c_5}{\tilde R{~\rm sh}^2\left(\sqrt{c_5}x\right)}.
\end{equation}
At last, in the case of $c<0$ solution (\ref{Rconst2c}) takes the form
\begin{equation} \label{Rconst2e}
z=\mp\frac{2}{\sqrt{-c}}\arcsin\left(\frac{\alpha\sqrt{-c}}{\sqrt{\tilde R}}\right)
\end{equation}
with
\begin{equation} \label{Rconst2f}
\alpha=\frac{\sqrt{\tilde R}}{\sqrt{-c}}\sin|\sqrt{-c}z/2|=e^{-y/2}.
\end{equation}
\begin{equation} \label{Rconst2g}
H=e^y=\frac{-c}{\tilde R~\sin^2\left(\sqrt{-c}z/2\right)}
=\frac{-c}{\tilde R~\sin^2\left(\sqrt{-c}x\right)}.
\end{equation}
Respectively, in the case $\tilde R<0$ we have $z=u+v$, $w_{uv}=w_{zz}$. The Liouville equation takes the form
\begin{equation}
\omega_{,zz}=-\frac{\tilde R}{4}e^{2\omega}=\frac{|\tilde R|}{4}e^{2\omega}.
\end{equation}
All the preceding solutions are reproduced by substitution $\tilde R \to |\tilde R|$ and
$x \tilde R \to t$.

We see that, up to the overall conformal factor, all the solutions with constant two-dimensional curvature $\tilde R$ can be cast in the Robertson-Walker form. And for $\tilde R > 0$ they represent the isotropic and homogeneous universes with either positive, negative or zero spatial curvature. If $\tilde R = + 2$, these universes are empty and with zero Weyl tensor. It is such a solution with positive spatial curvature that we called ``the gravitational bubble''. The case $\tilde R < 0$ differs from the above one in that the Robertson-Walker radial coordinate becomes time-like. Thus, we obtained the pure vacuum curved space-times (without any material sources, including the cosmological constant) what is absolutely impossible in General Relativity. Such a phenomenon makes is easier to create the universe from ``nothing''.

Note also, that the very existence of two different spherically symmetric vacua manifests the spontaneous symmetry breaking, because the vacuum with $\tilde R=2$ is isotropic, while that one with $\tilde R=-2$ --- is not. Surely, this symmetry breaking is not what one wants and needs (emergent General Relativity, the possibility to have the matter energy-momentum tensor with nonzero trace and so on), but still this provides us with some hopes.

\section{General structure of the energy-momentum tensor \\
	and vectorial equation}
\label{EnMomTensor}

In the preceding Section we found the non-vacuum solutions with constant two-dimensional curvature scalar $\tilde R$ for our specific choice of the spherically symmetric metrics,
\begin{equation}
\label{choice}
d s_4^2 = \tilde \gamma_{ik} d x^i d x^k - (d\theta ^2 + \sin^2 \theta d \phi^2) \, .
\end{equation}
The corresponding energy-momentum tensor equals
\begin{eqnarray}
&&\tilde T_{ik}=\alpha_0 \frac{\tilde R^2 - 4}{3} \tilde \gamma_{ik}, \quad \tilde T^2_2 = \tilde T^3_3 = - \frac{\alpha_0}{3} (\tilde R^2 - 4),
\nonumber \\
&&{\rm Tr} (\tilde T^{\mu}_{\nu}) = {\rm Tr} (\tilde T^i_k )+ 2 \tilde T^2_2 = \tilde T + 2 \tilde T^2_2 = 0.
\end{eqnarray}
In the present Section we are going to rewrite the tensorial spherically symmetric Bach equations in the more simple and, visually more convenient, vectorial form.

But, before doing this, some important notes are in order. This concerns the structure of the energy-momentum tensor in the conformal gravity. First (and this is the commonly mentioning fact), the energy-momentum tensor should be traceless. This follows immediately from the required conformal invariance of the matter part of the total action integral. Indeed, under the conformal variation of the metric tensor, $\delta g_{\mu \nu} = 2 e^{2 \omega} \tilde g_{\mu \nu} \delta \omega 2 g_{\mu \nu} \delta \omega \,$ $( \delta g^{\mu \nu} = - 2 g^{\mu \nu} \delta \omega)$,
\begin{eqnarray}
\label{teraceT}
0 &=& \delta S_{\rm matter} = \delta \int L_{\rm matter} \sqrt{- g} d x \stackrel{def}{\equiv} \frac{1}{2}\int T_{\mu \nu} \sqrt{- g} \, \delta g^{\mu \nu} d x \nonumber \\
&=& - \int T_{\mu \nu} g^{\mu \nu} \sqrt {-g} \,\delta \omega d x \; \; \; \Longrightarrow  {\rm Tr} (T^{\mu}_{\nu}) = 0.
\end{eqnarray}
Second, let us compare two energy-momentum tensors for two metrics related by the conformal transformation. For this, we should consider another type of variation which does not affect the conformal factor, namely, $\delta g^{\mu \nu} = e^{- 2 \omega} \delta \tilde g^{\mu \nu}$. Again,
\begin{eqnarray}
\label{conffac}
0 &=& \delta S_{\rm matter} = \delta \int L_{\rm matter} \sqrt{- g} d x  \stackrel{def}{\equiv} \frac{1}{2} \int T_{\mu \nu} \sqrt{- g} \, \delta g^{\mu \nu} d x \nonumber \\
&=& \frac{1}{2} \int \tilde T_{\mu \nu} \sqrt{- \tilde g} \, \delta \tilde g^{\mu \nu} d x \,\,\, \Longrightarrow 
\sqrt{- \tilde g} T_{\mu \nu} \, \delta g^{\mu \nu} = \sqrt{- \tilde g} \tilde T_{\mu \nu}\, \delta \tilde g^{\mu\nu}.
\end{eqnarray}
Therefore,
\begin{equation}
\label{TtildeT}
T_{\mu \nu} = e^{- 2 \omega} \tilde T_{\mu \nu}, \quad
T^{\nu}_{\mu} = g^{\nu \lambda} T_{\mu \lambda} = e^{- 4 \omega} \tilde g^{\nu \lambda} \tilde T_{\mu \lambda} = e^{- 4 \omega} \tilde T^{\nu}_{\mu}.
\end{equation}
It can be shown (see Appendix A) that the same law holds for the Bach tensor as well,
\begin{equation}
\label{Bachconf}
B^{\nu}_{\mu} = e ^{- 4 \omega} \tilde B^{\nu}_{\mu} \, ,
\end{equation}
therefore, everything is self-consistent. By the way, for the general form of the spherical metrics, namely,
\begin{equation}
\label{genmetr}
d s_4^2 = \Phi^2 (x) (\tilde y_{ik} d x^i d x^k - (d \theta^2 + \sin^2 \theta d\varphi^2))
\end{equation}
where $\Phi (x)$ plays the role of the physical radius of the sphere, the structure of the energy-momentum tensor in the case of constant two-dimensional curvature scalar $\tilde R$, is
\begin{equation}
\label{alacoulomb}
T^{\nu}_{\mu} = \frac{\alpha_0}{3} \frac{\tilde R^2 - 4}{\Phi^4}\, diag (1, \, 1, \, -1, \, -1)
\end{equation}
 i.\,e., exactly the same as for the spherically symmetric Coulomb-like field.

Now, let us come to the point. We need only the two-dimensional part of the Bach equations
\begin{equation}
\label{Bachvect}
\tilde B_{ik} = \frac{1}{6} \left ( \tilde R^{|p}_{|p} \tilde \gamma_{ik} - \tilde R_{|ik} \right) + \frac{\tilde R^2 - 4}{24} \tilde \gamma_{ik} = \frac{1}{8 \alpha_0} \tilde T_{ik}
\end{equation}
written for our specific choice of the metrics (without radius). The equation for $\tilde B^2_2 \, (= \tilde B^3_3)$ is the algebraic consequence of the above ones, since the bach tensor is traceless. We see, that the two-dimensional curvature scalar $\tilde R$ plays the very important role. Geometrically, the world-lines $\tilde R = const$ could be time-like, space-like, or null. This is determined by the (pseudo-Euclidian) square of the normal vector, we will denote it $\tilde \Delta$:
\begin{equation}
\label{tdelta}
\tilde \Delta = \tilde \gamma^{ik} \tilde R_{, i} \tilde R_{,k},  \quad (0=d\tilde R =\tilde R_{, i} d x^i).
\end{equation}
So, $\tilde \Delta > 0$ for space-like, $\tilde \Delta < 0$ for time-like and $\tilde \Delta = 0$ for null world-lines $\tilde R = const$. Taking trace of the Eq.(\ref {Bachvect}) and solving it for $\tilde R^{|p}_{|p}$ we arrive at
\begin{equation}
\label{traceR}
\tilde R^{|k}_{|i} + \frac{\tilde R^2 - 4}{4} \delta^k_i  = \frac{3}{4 \alpha_0} \left( \tilde T \, \delta^k_i - \tilde T^k_i \right) \, ,
\end{equation}
where $\tilde T = \tilde \gamma^{ik} \tilde T_{ik}$. The convolution with the co-vector $\tilde R_{|i} = \tilde R_{,i}$ gives us
\begin{equation}
\label{almost}
\tilde R^{|k}_{|i} \tilde R_{|k} + \frac{1}{4} (\tilde R^2 - 4) \tilde r_{|i} = \frac{3}{4 \alpha_0} \left( \tilde T \tilde R_{|i} - \tilde T^k_i \tilde R_{|k} \right) \, .
\end{equation}
The first term can be expressed in terms of $\tilde \Delta$:
\begin{equation}
\label{deltahelp}
\tilde \Delta_{,i} = \tilde \Delta_{|i} = 2 \tilde R^{|k}_{|i} \tilde R_{|k}\,,
\end{equation}
and we get the desired vectorial equation,
\begin{equation}
\label{vectorial}
\left(2 \tilde \Delta + \frac{1}{3} (\tilde R^3 - 12 \tilde R) \right)_{,i} = \frac{3}{\alpha_0} \left(\tilde T \tilde R_{,i} - \tilde T^k_i  \tilde R_{,k}\right) \, .
\end{equation}
It is not difficult to show that the integrability condition is provided by the continuity equation for the energy-momentum tensor.

As an example, let us apply this vectorial equation to the situation when $\tilde R$, as well as the physical radius $\Phi = r$ can be used as the spatial coordinate. Then, $\tilde \Delta = \gamma^{11} < 0$, and
\begin{equation}
\label{example}
- \gamma^{11} = \frac{1}{6} \left(\tilde R^3 - 12 \tilde R \right) - \frac{3}{2 \alpha_0} \int \tilde T^0_0 d \tilde R \, .
\end{equation}
Assuming now that $\tilde R = \tilde R (r)$ and remembering that for the Mannheim--Kazanas solution we had $\frac{d \tilde R}{d r} = \frac{C_1}{r^2}$, the term with the integral can be rewritten as follows:
\begin{equation}
\label{massterm}
\int\limits_0^{r_0} \tilde T^0_0 d \tilde R = \int\limits_0^{r_0} r^4 T^0_0 \frac{d \tilde R}{d r} d r = C_1 \int\limits_0^{r_0} T^0_0 r^2 d r \, ,
\end{equation}
what is proportional to the total mass, including the gravitational mass defect, inside the sphere of the radius $r_0$, as it was defined in \cite{LL2}.

\section{Solution \`a la Vaidya}
\label{Vaidya}

The Vaidya solution is the solution of the Einstein equations for the gravitating spherically symmetric radiation, radially outgoing or ingoing. We would like to find its counterpart in the discussed conformal gravity. To be specific, we consider in details the case of the outgoing rays.

As before, we should, first, choose the form of the 2-dimensional metrics. Evidently, the most suitable coordinates now are the null coordinate $u$ (``retarded time'') and the 2-dimensional  curvature scalar $\tilde R$:
\begin{equation}
ds_2^2=Adu^2+2Bdud\tilde R.
\end{equation}
We will use the index ``0'' for $u$ and ``1'' for $\tilde R$. Then metric coefficients $\gamma_{00}=A$, $\gamma_{01}=\gamma_{10}=B$, $\gamma_{11}=0$, respectively, for the inverse metric, $\gamma^{00}=0$, $\gamma^{01}=\gamma^{10}=1/B$, $\gamma_{11}=-A/B^2$. The energy-momentum tensor for the radiation has the form
\begin{equation}
T_{ik}=l_il_k, \quad (i,k=0,1),
\end{equation}
where $l^i$ is the null vector: $\gamma_{ik}l^il^k=0$. Hence $A(l^0)^2+2Bl^0l^1=0$. For the outgoing rays, $l^0=0$ and $l_0=Bl^1$, $l_1=0$, so
\begin{equation}
T_{00}=B(l^1)^2, \quad T_{01}=T_{10}=T_{11}=0.
\end{equation}
The corresponding Bach equations are
\begin{eqnarray}
B_{00} &=& \frac{1}{6}\!\left[A\frac{\tilde R^2-4}{4}\!
+\!\frac{1}{B}\left(\frac{A'}{2}\!-\!\frac{AB'}{B}\right)\!\left(1
\!-\!\frac{A}{B^2}\right)\!\right] 
=-\frac{1}{8\,\alpha}T_{00},
\\
B_{01} &=& \frac{B}{6}\left(\frac{A'B'}{B^3}-\frac{A'}{2B^2}+\frac{\tilde R^2-4}{4}=0
\right)\!, \\
B_{11} &=& \frac{1}{6}\frac{B'}{B}=0.
\label{B01}
\end{eqnarray}
From the last equation we have $B=B(u)$ and by rescaling the null coordinate, we can make $B=\pm1$. By taking limit of the well known Schwarzschild solution, we conclude that for the outgoing rays there should be $B=1$. From (\ref{B01}) for the metric coefficient $A$ it is now easy to get
\begin{equation}
A=\frac{1}{6}[\tilde R^3 -12\tilde R+C_0(u)],
\end{equation}
where the integration ``constant'' $C_0(u)$ is an arbitrary function of $u$. Thus
\begin{equation}
ds_2^2=\frac{1}{6}[\tilde R^3 -12\tilde R+C_0(u)]du^2+2dud\tilde R.
\end{equation}
We see that the only difference from the vacuum solution is that the constant of integration $C_0$ becomes now the function of the retarded time (outgoing null coordinate), what confirms our interpretation of the total mass of the gravitating source as the part of $C_0$.  Of course, to get the general solution, we should add the metrics of the 2-dimensional unit sphere, and then multiply all this by the physical radius squared.

\section{Electrovacuum solution}

It was already mentioned that the energy-momentum tensor for the spherically symmetric Coulomb field equals
\begin{equation}
T_\mu^\nu=\frac{e^2}{8\pi\Phi^4}{\rm diag}(1,1,-1,-1).
\end{equation}
Consequently, its two-dimensional truncated counterpart is simply
\begin{equation}
\tilde T_i^k=\frac{e^2}{8\pi}\delta_i^k,
\end{equation}
where $\delta_i^k$ is the unit tensor (Kronecker symbol). Substituting this into our vectorial equation, one gets immediately
\begin{equation}
A=\frac{1}{6}\left[\tilde R^3 -\left(12+\frac{9e^2}{4\pi\alpha_0}\right)\tilde R+const\right].
\end{equation}
The structure of the metrics is the same as in the vacuum case. Note, that now it is impossible to recover the famous Reissner--Nordstrom solution of General-Relativity for any choices of the constant of integration and the radius $\Phi(\tilde R)$.

\section{Conclusion}

Here we summarize the results obtained, and make some notes. The general structure of the spherically symmetric solutions in the Weyl conformal gravity is described. The corresponding Bach equation are derived for the special type of metrics, which can be considered as the representative of the general class. The complete set of the pure vacuum solutions is found. It consists of two classes.

The first one contains the solutions with constant two-dimensional curvature scalar of our specific metrics, and the representatives are the famous Robertson--Walker metrics. One of them we called the ``gravitational bubbles'', which is compact and with zero Weyl tensor. Thus, we obtained the pure vacuum curved space-times (without any material sources, including the cosmological constant) what is absolutely impossible in General Relativity. Such a phenomenon makes is easier to create the universe from ``nothing''. There are two different spherically symmetric vacua (up to the conformal transformation). This manifests the spontaneous symmetry breaking, because the vacuum with $\tilde R=2$ is isotropic, while that one with $\tilde R=-2$ --- is not. Surely, this symmetry breaking is not what one wants and needs (emergent General Relativity, the possibility to have the matter energy-momentum tensor with nonzero trace and so on), but still this provides us with some hopes.

The second class is more general, with varying curvature scalar. We found its representative as the one-parameter family. It appears that it can be conformally covered by the thee-parameter Mannheim--Kazanas solution. 

The difference between our solution and the Mannheim-Kazanas one is that the latter depends on three integration constants up to the overall arbitrary conformal factor, while ours has only one integration constant (and an arbitrary conformal factor). It is also shown that two extra constants appear in the process of partial restoration of the arbitrariness in the general form of the metrics. In this way our solution is more general and original. 

We also investigated the general structure of the energy-momentum tensor in the spherical conformal gravity and constructed the vectorial equation that reveals clearly the same features of non-vacuum solutions. Two of them, the metrics \`a la Vaidya and the electrovacuum metrics, are explicitly written. 

We showed that for the equations of motion to be conformally invariant it is sufficient to have the traceless energy-momentum tensor of the matter fields. Moreover, for two metrics that differ only by the conformal factor, the corresponding energy-momentum tensors differ by definite powers of this very conformal factor. In other words, the pure conformal freedom in choosing the metrics holds only for the vacuum space-times.

Some feature of the vacuum solution with varying two-dimensional truncated scalar curvature $\tilde R$ ( i.\,e., outside the material source) may allow us to explain the observed accelerated expansion in our universe. Indeed, let us assume that $\tilde R$ is increasing function of the radius (like in General Relativity) and put the observer at the center. Then, adding the ordinary (not exotic) matter layer by layer we will reduce more and more the constant $C_0$ in the solution. This corresponds to the greater and greater values of the effective cosmological constant outside these layers. Eventually one comes to the situation when attraction of the test bodies will be replaced by the repulsion. In a sense, this is the kinematic  effect, it depends on the position of central observer and the amount of matter between his (her) and the point of observation. Moreover, it does not mean that the universe as a whole is expanding with the acceleration.

We intend to apply and generalize the above formalism to the initial state of the universe and the subsequent particle creation. In a sense, it can be considered as the pre-inflationary stage.

\appendix
\section{Derivation of Bach equations}
\label{Bacheqapp}

The variation of the action (\ref{C2}) is
\begin{eqnarray}
\delta S&=& -\alpha_0\delta\int C_{\mu\nu\lambda\sigma}C_{\mu'\nu'\lambda'\sigma'}g^{\mu\mu'}g^{\nu\nu'}
g^{\lambda\lambda'}g^{\sigma\sigma'}\sqrt{-g}d^4x
\\
&=&-2\alpha_0\int C^{\mu\nu\lambda\sigma}(\delta C_{\mu\nu\lambda\sigma})\sqrt{-g}d^4x \label{term1}
-4\alpha_0\int C_{\varkappa\lambda\sigma\gamma}C_{\rho}^{\phantom{0}\lambda\sigma\gamma}\delta g^{\varkappa\rho}\sqrt{-g}d^4x \label{term2}
\\
&&-\alpha_0\int C^2(\delta\sqrt{-g})d^4x, \label{term3}
\end{eqnarray}
where we used the notation $C^2\equiv C^{\mu\nu\lambda\sigma}C_{\mu\nu\lambda\sigma}$.
To calculate (\ref{term1}) we write down $C_{\mu\nu\lambda\sigma}$ in terms of the Riemann and Ricci according to the definition (\ref{Weyltensor}). Then, equation(\ref{term1}) becomes
\begin{equation}
-2\alpha_0\int C^{\mu\nu\lambda\sigma}\left[\delta R_{\mu\nu\lambda\sigma}-2R_{\mu\lambda}(\delta g_{\nu\sigma})\right]\sqrt{-g}d^4x.
\label{term1-2}
\end{equation}
The major part of terms are canceled due to symmetries of the Weyl tensor $C^{\mu\nu\lambda\sigma}$. Now, writing $R_{\mu\nu\lambda\sigma}=g_{\mu\varkappa}R^{\varkappa}_{\phantom{0}\nu\lambda\sigma}$ we get
\begin{equation}
\delta R_{\mu\nu\lambda\sigma}=(\delta g_{\mu\varkappa})R^{\varkappa}_{\phantom{0}\nu\lambda\sigma}+g_{\mu\varkappa}(\delta R^{\varkappa}_{\phantom{0}\nu\lambda\sigma}).
\end{equation}
In the following we use equations (16.3) and (16.4) from \cite{DeWitt}:
\begin{equation}
\delta \Gamma^{\lambda}_{\mu\sigma}=\frac{1}{2}g^{\lambda\nu}\left[\left(\delta g_{\nu\mu}\right)_{;\sigma}+\left(\delta g_{\nu\sigma}\right)_{;\mu}-\left(\delta g_{\mu\sigma}\right)_{;\nu}\right],
\end{equation}
\begin{equation}
\delta R^{\varkappa}_{\phantom{0}\nu\lambda\sigma}=\left(\delta \Gamma^{\varkappa}_{\nu\sigma}\right)_{;\lambda}-\left(\delta \Gamma^{\varkappa}_{\nu\lambda}\right)_{;\sigma}.
\end{equation}
As the next step, we integrate the (\ref{term1}) by parts and use the Gauss theorem. When omitting the surface terms, (\ref{term1}) is reduced to
\begin{equation}
-2\alpha_0\int\! \left[C^{\mu\varkappa\lambda\sigma}R^{\nu}_{\phantom{0}\varkappa\lambda\sigma}
+2C^{\mu\sigma\lambda\nu}_{\phantom{0}\phantom{0}\phantom{0}\phantom{0};\lambda;\sigma}
-2C^{\sigma\nu\lambda\mu}R_{\sigma\lambda}
\right]\delta g_{\mu\nu}\sqrt{-g}\,d^4x. \label{term1-3}
\end{equation}
By using formulas  $\delta g^{\varkappa\rho}=-g^{\varkappa\mu}g^{\rho\nu}\delta g_{\mu\nu}$ and $\delta\sqrt{-g}=\sqrt{-g}g^{\mu\nu}\delta g_{\mu\nu}/2$, we are able to write the terms (\ref{term2}) and (\ref{term3}), respectively, in the forms
\begin{equation}
4\alpha_0\int C^{\mu}_{\phantom{0}\lambda\sigma\gamma}C^{\nu\lambda\sigma\gamma}\delta g_{\mu\nu}\sqrt{-g}d^4x\label{term2-3}
\end{equation}
and
\begin{equation}
-\frac{1}{2}\alpha_0\int C^2g^{\mu\nu}\delta g_{\mu\nu}\sqrt{-g}d^4x.\label{term3-3}
\end{equation}
The first term in the square brackets in (\ref{term1-3}) we transform with the help of the relation \begin{equation}
C^{\mu\varkappa\lambda\sigma}C^{\nu}_{\phantom{0}\varkappa\lambda\sigma}
=C^{\mu\varkappa\lambda\sigma}R^{\nu}_{\phantom{0}\varkappa\lambda\sigma}
+C^{\mu\varkappa\lambda\nu}R_{\varkappa\lambda},
\end{equation}
which follows from the definition of the Weyl tensor (\ref{Weyltensor}). Then, from the sum of (\ref{term1-3}), (\ref{term2-3}) and (\ref{term3-3}) we obtain the field equations in the form
\begin{equation}
C^{\mu\sigma\nu\lambda}_{\phantom{0}\phantom{0}\phantom{0}\phantom{0};\lambda;\sigma}
+\frac{1}{2}C^{\mu\lambda\nu\sigma}R_{\lambda\sigma}
+\frac{1}{2}C^{\mu\varkappa\lambda\sigma}C^{\nu}_{\phantom{0}\varkappa\lambda\sigma}
-\frac{1}{8}C^2g^{\mu\nu}=0.
\end{equation}
Note that the last two terms are canceled due to relation
\begin{equation}
C^{\mu\varkappa\lambda\sigma}C^{\nu}_{\phantom{0}\varkappa\lambda\sigma}
=\frac{1}{4}C^2g^{\mu\nu},
\end{equation}
which is an algebraic relation (16.35) from \cite{DeWitt}, rewritten in terms of the Weyl tensor. Here it must be taken in mind that both in the metric and in the Weyl tensor in \cite{DeWitt} there is a multiplier $(-1)$ in comparison with the definitions of \cite{LL2}, which are used here.
Finally, the vacuum field equations (the Bach equations) take the form
\begin{equation}
C^{\mu\sigma\nu\lambda}_{\phantom{0}\phantom{0}\phantom{0}\phantom{0};\lambda;\sigma}
+\frac{1}{2}C^{\mu\lambda\nu\sigma}R_{\lambda\sigma}=0.
\label{bachvac}
\end{equation}
The 2nd term in (\ref{bachvac}) is symmetric in $\mu$ and $\nu$ due to the symmetries of $C^{\mu\lambda\nu\sigma}$ and $R_{\lambda\sigma}$. To show the symmetry of the 1st term one can write
\begin{equation}
C^{\mu\sigma\lambda\nu}_{\phantom{0}\phantom{0}\phantom{0}\phantom{0};\rho;\varkappa}
-C^{\mu\sigma\lambda\nu}_{\phantom{0}\phantom{0}\phantom{0}\phantom{0};\varkappa;\rho}
=-C^{\gamma\sigma\lambda\nu}R^{\mu}_{\phantom{0}\gamma\rho\varkappa}+...
\end{equation}
and obtain that $C^{\mu\sigma\nu\lambda}_{\phantom{0}\phantom{0}\phantom{0}\phantom{0};\lambda;\sigma}
=C^{\mu\sigma\nu\lambda}_{\phantom{0}\phantom{0}\phantom{0}\phantom{0};\sigma;\lambda}$ identically.

If there is a matter with the energy-momentum tensor $T^{\mu\nu}$, then the matter variation is \cite{LL2}
\begin{equation}
\delta S_m=-\frac{1}{2}\int T^{\mu\nu}(\delta g_{\mu\nu})\sqrt{-g}d^4x,
\end{equation}
and the field equations become
\begin{equation}
C^{\mu\sigma\nu\lambda}_{\phantom{0}\phantom{0}\phantom{0}\phantom{0};\lambda;\sigma}
+\frac{1}{2}C^{\mu\lambda\nu\sigma}R_{\lambda\sigma}
=\frac{1}{8\alpha_0}T^{\mu\nu}.
\end{equation}
If the action (\ref{C2}) contains the variable coefficient $\lambda(x)$ instead of the constant $\alpha_0$, the corresponding field equations take the form
\begin{equation}
\left(\lambda(x)C^{\mu\sigma\nu\lambda}_{\phantom{0}\phantom{0}\phantom{0}\phantom{0}}\right)_{;\lambda;\sigma}
+\frac{1}{2}\lambda(x)C^{\mu\lambda\nu\sigma}R_{\lambda\sigma}
=\frac{1}{8}T^{\mu\nu}.
\end{equation}

Note that the conformal invariance of the action (\ref{C2}) does not mean the conformal invariance of the Bach tensor
\begin{equation}
B^\mu_{\phantom{0}\nu}\equiv C^{\mu\phantom{0}\phantom{0}\phantom{0};\lambda;\sigma}_{\phantom{0}\sigma\nu\lambda}
+\frac{1}{2}C^{\mu}_{\phantom{0}\lambda\nu\sigma}R^{\lambda\sigma}.
\end{equation}
Under the conformal transformation
\begin{equation}
\tilde g_{\mu\nu}=e^{2\omega}g_{\mu\nu}
\end{equation}
the Bach tensor transforms as
\begin{equation}
\tilde B^\mu_{\phantom{0}\nu}=e^{-4\omega}B^\mu_{\phantom{0}\nu}.
\end{equation}
So, the conformal transformation of the metrics must be accompanied by the corresponding transformation of the energy-momentum tensor.

\acknowledgments

The reported study was partially supported by RFBR, research project No. 13-02-00257~a.

\end{document}